\documentclass{agujournal}

\journalname{JGR-Oceans}

\usepackage{graphicx}
\usepackage{color}
\usepackage{amsmath,amsfonts,amssymb}

\newcommand{\zint}{\int_{-h}^0}

\newcommand{\be}{\begin{equation}}
\newcommand{\ee}{\end{equation}}

\newcommand{\bk}{\mathbf{k}}
\newcommand{\bU}{\mathbf{U}}
\newcommand{\kU}{\bk\cdot\bU}
\newcommand{\kdU}{\bk\cdot\Delta\bU}

\newcommand{\rmd}{\mathrm{d}}

\newcommand{\tc}{\tilde{c}}
\newcommand{\tU}{\tilde{U}}

\begin{document}

\title{Approximate dispersion relations for waves on 
arbitrary shear flows}

\authors{S. {\AA}. Ellingsen\affil{1} and Y. Li\affil{1}\thanks{S.\AA.E.\ and Y.L.\ are to be considered joint first authors, ordered alphabetically.}}

\affiliation{1}{Department of Energy and Process Engineering, Norwegian University of Science and Technology, Trondheim, Norway}

\correspondingauthor{Y. Li}{yan.li@ntnu.no}


\begin{keypoints}
\item 3D approximate dispersion relation for surface waves on shear current whose magnitude and direction vary arbitrarily with depth
\item The approximation is robust and accurate up to 2nd order with explicit estimates of error and criteria of applicability
\item Compares favorably to approximation by Kirby \& Chen [1989] whose properties and error estimate are derived for the first time
\end{keypoints}

%
%


\begin{abstract}
  An approximate dispersion relation is derived and presented for linear surface waves atop a shear current whose magnitude and direction can vary arbitrarily with depth. The approximation, derived to first order of deviation from potential flow, is shown to produce good approximations at all wavelengths for a wide range of naturally occuring shear flows as well as widely used model flows. 
  The relation reduces 
  in many cases
  to a 3D generalization of the much used approximation by 
  Skop [1987], developed further by 
  Kirby \& Chen [1989], but is shown to be more robust, succeeding in situations where the Kirby \& Chen model fails. The two approximations incur the same numerical cost and difficulty. 
  While the Kirby \& Chen approximation is excellent for a wide range of currents, the exact criteria for its applicability have not been known. 
  We explain 
  the apparently serendipitous success of the latter
  and derive proper conditions of applicability for both approximate dispersion relations.
  Our new model has a greater range of applicability. 
  A second order approximation is also derived. It greatly improves accuracy, which is shown to be important in difficult cases. It has an advantage over the corresponding 2nd order expression proposed by Kirby \& Chen that its criterion of accuracy is explicitly known, which is not currently the case for the latter to our knowledge. Our 2nd order term is also arguably significantly simpler to implement, and more physically transparent, than its sibling due to Kirby \& Chen. 
\end{abstract}

\section{Introduction}

Knowledge of the properties of surface waves in the presence of currents is key to understanding, measuring and monitoring a range of processes in the oceans. Exchange of energy, mass and momentum between ocean and atmosphere is pivotal to predicting climate change, and local wave-current flow conditions can strongly affect the spread of nutrients as well as pollutants. Moreover, a broad range of fixed, floating or moored installations and vessels are affected by loads from waves and currents in combination; pertinent examples include floating oil booms which may fail under waves and currents, aquaculture farms in exposed locations, vessels, robots operating near the surface and even biolocomotion.  Measuring currents from wave observations, e.g.\ using radar \citep{lund15} is a favoured technique for obtaining flow field information towards such ends, but requires knowledge of how sheared currents affect wave dispersion. 
The effect of shear on waves is now included in widely employed oceanographic models such as Delft-3D \citep{elias12} and ROMS, used for example in the coupled COAWST model \citep{kumar11,kumar12}, and is recently employed for currents measurements using x--band radar observation of waves \citep{lund15,campana17}. A potentially important piece of progress towards incorporating arbitrary shear currents in oceanographic models was the recent derivation of an explicit wave action conservation equation on such currents by \cite{quinn17}. They conclude that the neglect of shear in wave modelling in realistic ocean conditions can lead to 
significant
errors.

In the coastal zone in particular, currents are often strongly sheared, and may change direction beneath the surface, for example when wind is blowing across a tidal current. In such conditions a fully 3D approximate dispersion relation is required to correctly analyse the dispersion properties of surface waves, i.e., how the phase and group velocities of waves vary as a function of wavelength propagation direction and the shape and magnitude of the subsurface current. We present herein an approximation valid for an arbitrary horizontal velocity field $ \mathbf{U} (z)$ whose magnitude and direction may change as a function of $z$, providing good estimates of $c( \mathbf{k} )$ for all $ \mathbf{k} $ for typically occurring velocity profiles. This main result is found in equation \eqref{cEL}. 

Under many, but 
far from
all, flow conditions of practical importance, the model coincides with a 3D generalization of the much used approximation of \citet{kirby89} to leading order in an expansion in terms of ``deviation from potentiality'', and involves the same calculational cost and complexity as the latter. However, several examples of realistic flows are presented where the Kirby \& Chen model fails while our new one performs well or at least remains reasonable, demonstrating the improved robustness of our model. We also show that the criterion for our model to produce accurate approximations is less stringent than that of the Kirby \& Chen approximation. Our derivations automatically produces a 2nd order accurate approximation
--- equation \eqref{c2nd} --- 
which is shown to be robust, and is arguably simpler and more transparent than the corresponding 2nd order approximation due to Kirby \& Chen. 

Although several approximations exist for the 2D and (with straightforward generalization) 3D flow situation as reviewed below, some of which are excellent in most practical cases, there is a lack of understanding of their conditions of applicability and estimated error. We here rectify this situation by providing a thorough discussion of the accuracy and criteria of applicability of our approximation model as well as that of Kirby \& Chen and a short-wavelength approximation due to \citet{shrira93}. 

While work on waves on arbitrary shear currents is relatively sparse, a large body of literature exists on waves propagating on linearly varying currents in 2D. We make no attempt to review this very large literature, but mention but a few important papers. 
Key non--linear treatments are those of \citet{simmen85}, \citet{telesdasilva88}. 
Of particular interest is the numerical scheme for fully nonlinear stationary and solitary waves on arbitrary 2D shear currents due to \citet{dalrymple77}, and followed by further numerical work on this question \citep{ko08a,ko08b,nwogu09,moreira15}. The present effort is restricted to linear waves; a comparison of the effect on wave speed of shear vs that of non-linearity is an important topic for the future; refer to \cite{swan00} for some results 
up to second order in wave steepness.

\subsection{Existing approximation methods}

Probably the most successful and widely used approximate dispersion relation is that due to \citet{kirby89}, which in turn was a direct generalisation of previous models by \citet{stewart74} and \citet{skop87}. Their approximation includes a 2nd order correction term, but we shall mostly be concerned with their first order correction, 
identical to that of \citet{skop87}. 
By the Kirby \& Chen approximation the first order expression is meant, unless otherwise stated. We will work with a direct generalisation of Kirby \& Chen's model to 3D (3DKC). The success of Kirby \& Chen's formula is likely in part to be due to the fact that, apparently serendipitously, the approximation it produces is often excellent even when the assumptions from which it was derived are strongly violated. In our analysis in the following we are able to explain this fortunate circumstance, which we believe has not previously been fully understood. 

Among analytical dispersion relation approximations for arbitrary shear currents, that of \citet{shrira93} is the only one to our knowledge explicitly derived for a fully 3D configuration such as we consider herein, where the current may vary in both direction and magnitude with depth. Generalization of the Kirby \& Chen approximation to 3D, however, is straightforward. Shrira's relation is of limited use for oceanographic purposes because it is accurate only for very short waves.
It is reviewed further in appendix \ref{app:shrira}.

An altogether different approach is to approximate the smooth velocity profile by a series of $N$ piecewise segments; we refer to this approach as the $N$-layer model. This approach, which goes back a long time \citep[e.g.][]{raileigh1892,thompson49,taylor55}, has the virtue that for a given wavevector $ \mathbf{k} $ it converges as $N^{-2}$ to the exact value of $c( \mathbf{k} )$ as the number of layers grows \citep{zhang05}, making it a useful comparison for the models developed here, 
and is preferable in cases where close control of errors, or particularly high accuracy, are required. 
The $N$-layer model was recently implemented by \citet{smeltzer16} in 3D, albeit with unidirectional shear current. The $N$-layer approach is suitable when high and closely controlled accuracy for all wavelengths and directions is required.
While numerically cheap, it remains more cumbersome to implement than analytical approximations due to the need to eliminate additional, spurious solutions \citep{smeltzer16}.

Some further approaches to approximating wave dispersion on curved velocity profiles should be mentioned, with no ambition of completeness. \citet{zhang05} presents an analytical approximation which amounts to a matching of the short-wavelength model of Shrira to the 3DKC model at longer wavelengths. Solution schemes have also been developed for specific velocity profiles; exponential current profiles modelling wind-driven currents were studied by \citet{abdullah49} and the $1/7$ law profile by \citet{hunt55,fenton73}. Several examples are worked out with a method developed by \citet{karageorgis12}. The case of shallow waters compared to wavelength was considered by \citet{burns53}. Numerical approaches, using shooting methods, have been used e.g.\ by \citet{fenton73, dong12}.

\subsection{Approach}

The widely used approximation by \citet{kirby89}, like its predecessors by \citet{stewart74} and \citet{skop87}, all proceed by performing a formal expansion in orders of a dimensionless parameter which is assumed small \textit{a priori}. Our present approach is somewhat different, and is inspired by that of \cite{shrira93}. We adopt a `near--potentiality' assumption, that the wave--induced fluid motion does not differ greatly from that due to a flow with linearly varying velocity profile. The formal procedure is to isolate the effect of the curvature of the velocity profile in a single term in an implicit non--closed dispersion relation, and combine it with an iterative solution to the Rayleigh equation by a method of dominant balances assuming the term resulting from curvature to be dominated. Like \citet{shrira93} a `true' corresponding small parameter comes out as a \emph{result} rather than an initial assumption, and may be interpreted as a suitably depth--averaged measure of the shear-profile's curvature.

\subsection{Outline}

In the next section our approximate dispersion relation, equation \eqref{cEL}, is presented and analysed. We 
analyse the question of the range of applicability of our perturbation scheme, providing order of magnitude estimates for the error. The relation to the 3DKC is discussed, including the conditions under which the two coincide to leading order. Corresponding approximations for group velocity are also presented. A brief discussion of critical layers is given in Section \ref{sec:critlayers}. In Section \ref{sec:numerics} we test our model for a range of realistic shear profiles, both unidirectional and varying in direction as a function of depth, including particular cases where the 3DKC breaks down. Summary and conclusions are found in Section \ref{sec:conclusions}, and some further information and details may be found in appendices.
%

\section{An approximate relation for 3D wave-shear current dispersion} \label{sec:app}

\begin{figure}
  \begin{center}\includegraphics[width=0.7\textwidth]{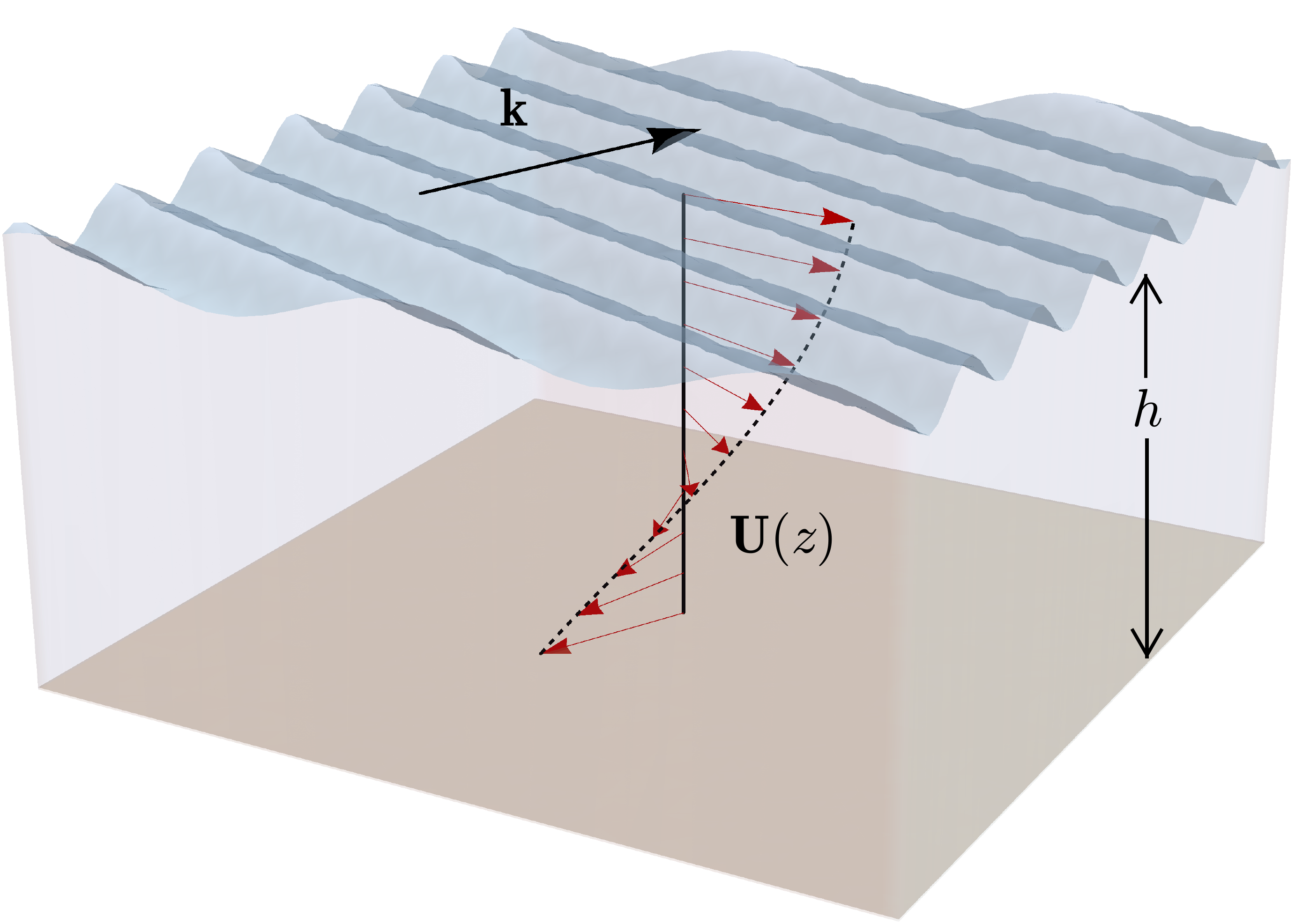}\end{center}
  \caption{The geometry: a plane wave of wave vector $ \mathbf{k} $ travels atop a horizontal shear current whose magnitude and direction may vary with depth.}
  \label{fig:geom}
\end{figure}

We consider the system shown in Fig.~\ref{fig:geom}. A horizontal current $ \mathbf{U} (z)$ whose direction and magnitude may vary with depth, running over a flat sea bed of depth $h$ and with a free surface which is at $z=0$ when undisturbed. 
Consider a plane wave with wave vector $ \mathbf{k} =[k_x,k_y]$ propagating atop the current. The wave creates a disturbance of velocity and pressure fields. We work within linear wave theory, hence all equations of motion and boundary conditions are linearized with respect to perturbations due to the wave motion. All perturbations are understood to be proportional to $\exp[\mathrm{i} \mathbf{k} \cdot\mathbf{x}-\mathrm{i} kc( \mathbf{k} ) t]$; here $\mathbf{x}=[x,y]$ is the position in the horizontal plane, $c$ is the phase velocity in direction $ \mathbf{k} $, $k=| \mathbf{k} |$, and $t$ is time. We neglect surface tension and viscosity, so the flow is governed by the Euler equation, which reduces to the following boundary value problem for the vertical velocity $w(z)$ only \citep[e.g.][]{shrira93},
\begin{subequations}\label{boundary}
\begin{eqnarray}
  w''(z)-k^2 w(z) &=& \frac{ \mathbf{k} \cdot \mathbf{U} ''(z)}{ \mathbf{k} \cdot \mathbf{U} (z)-kc}w(z);\label{rayleigh} \\
  ( \mathbf{k} \cdot \mathbf{U} _0-kc)^2 w'(0)-[ \mathbf{k} \cdot \mathbf{U} '_0( \mathbf{k} \cdot \mathbf{U} _0-kc)+gk^2]w(0)&=& 0;\label{surfBC}\\
  w(-h)&=&0\label{bottBC}.
\end{eqnarray}
\end{subequations}
Equation \eqref{rayleigh} is called the Rayleigh equation (or inviscid Orr--Sommerfeld equation), and equations \eqref{surfBC} and \eqref{bottBC} are the appropriate boundary conditions at the free surface and bottom, respectively. We defined $ \mathbf{U} _0= \mathbf{U} (0)$ and $ \mathbf{U} '_0= \mathbf{U} '(0)$,  and $g$ is gravitational acceleration. For future reference we define velocities relative to the surface
\begin{equation} 
  \tilde{c} = c -  \mathbf{k} \cdot \mathbf{U} _0/k; ~~ 
  \Delta \mathbf{U}(z)  =  \mathbf{U}(z)  -  \mathbf{U} _0.
\end{equation} 
Here $\tilde{c}$ is the intrinsic phase velocity.

Our task in this section will be to derive, 
discuss and test the approximate dispersion relation $c( \mathbf{k} )\approx c_\approx( \mathbf{k} )$ 
which is a main result and is given in Eq.~\eqref{cEL}.

Inspired by \citet{shrira93} we make a 'near--potentiality' assumption which entails that the effect of the current on the dispersion relation differs only moderately from the explicitly solvable case of a linearly varying $\bU(z)$. On the other hand, the shear itself need not be small. 
(Note that the term `near--potentiality' is a slight misnomer since a wave propagating at an oblique angle with such a current in fact has vorticity and is not as such `potential'; see \citet{ellingsen16}. Potential theory can be employed to waves on linearly depth--dependent currents only in strictly 2--dimensional flow \citep[e.g.][]{ellingsen14b}.)

\subsection{Re-casting the equations of motion}

We proceed in two steps. First, the boundary value problem \eqref{boundary} is written in an alternative way; combining the boundary conditions and Rayleigh equation we find an implicit dispersion relation \eqref{fulldispAlt} with both $\tc$ and $w$ unknown, where the effect of the depth--averaged shear is separated from that due to higher derivatives of $\bU(z)$. Approximations of $w$ are inserted into this equation, found by iterative solutions of \eqref{rayleigh} presuming the right--hand side to be a dominated term. In Section \ref{sec:approx} we will term the approximate dispersion relation accounting only for depth--averaged shear the `first order' approximation, and the improved approximation from including the term in \eqref{fulldispAlt} due to curvature as well as the second--level approximation for $w$, we term the `second order' approximation of $\tc$.

A fruitful starting point is to frame the boundary value problem \eqref{boundary} in the form,
\begin{equation} \label{fulldisp}
   \left[1+   I(\tc)   \right]k^2\tilde{c}^2
   + \tilde{c} \mathbf{k} \cdot \mathbf{U} '_0\tanh kh - k^2c_0^2 = 0
\end{equation} 
with
\begin{equation} \label{Iexp}
  I(\tc) = \zint \frac{ \mathbf{k} \cdot \mathbf{U} ''(z)w(z)\sinh k(z+h)}{kw(0)[ \mathbf{k} \cdot\Delta \mathbf{U} (z)-k\tilde{c}]\cosh kh}\mathrm{d} z 
\end{equation} 
and $c_0 = \sqrt{(g/k)\tanh kh}$. This is obtained by multiplying \eqref{rayleigh} by $\sinh k(z+h)$, integrating with respect to $z$, and inserting boundary conditions \eqref{surfBC} and \eqref{bottBC}. This very useful relation informs most of our analysis in this paper.
In itself the relation \eqref{fulldisp} is not closed since both $w$ and $\tilde{c}$ are unknown.

Eq.~\eqref{fulldisp} appears to isolate the effect of the curvature $\bU''$ in the term $I$, but this is deceptive: there is still an explicit reference to the surface shear $\bU'_0$ which may also be re-written as a depth--integrated curvature. The appearence of $\bU'_0$ is moreover undesirable: long waves in particular must be expected to be influenced by the suitably depth--averaged shear, wheras the value at the surface is comparatively irrelevant. 

To transform \eqref{fulldisp} into the desired form we write the integral $I(\tc)$ as
\begin{equation} \label{Iexp}
  I(\tc) 
  \equiv I_0 + \frac{c_0^2}{\tilde{c}^2}\varepsilon I(\tilde{c}) =  \frac{c_0}\tc \left(K  - \frac{\kU'_0}{k^2 c_0}\tanh kh\right)+\frac{c_0^2}{\tilde{c}^2}\varepsilon I(\tilde{c}),
\end{equation} 
with 
\begin{eqnarray}
  \varepsilon I(\tilde{c})&=&\frac{\tilde{c}}{c_0^2}\zint \frac{( \mathbf{k} \cdot \mathbf{U} '')( \mathbf{k} \cdot\Delta \mathbf{U} )w(z)\sinh k(z+h) }{k^2( \mathbf{k} \cdot\Delta \mathbf{U} -k\tilde{c})w(0)\cosh kh}\mathrm{d} z;\label{eI} \\  
  K &=& \zint \frac{ \mathbf{k} \cdot \mathbf{U} '}{k^2c_0\cosh kh}\frac{\mathrm{d}}{\mathrm{d} \zeta}\left[\frac{w(\zeta)}{w(0)}\sinh k(\zeta+h)\right]_{\zeta=z}\mathrm{d} z.\label{K}
\end{eqnarray}
Here $I_0$ is obtained from $I$ by setting $\Delta \mathbf{U} =0$, and the last form is obtained by partial integration of $I_0$.
Inserting into \eqref{fulldisp} gives
\begin{equation} \label{fulldispAlt}
  \tilde{c}^2+c_0\tilde{c} K+c_0^2\varepsilon I(\tilde{c})-c_0^2 = 0
\end{equation} 
which no longer contains an explicit reference to the surface. Instead the effect of the shear is contained in $K$, and the effect of the curvature shared between $K$ and $\varepsilon I$.

Equation \eqref{fulldispAlt} is exact within linear wave theory, and we now begin making approximations. 
To wit we wish to isolate the part of $K$ which depends on the shear but not the curvature, plus a curvature--related correction. 
We adopt an iterative approach rather than a formal expansion in an explicit parameter as done by \cite{kirby89} and predecessors. Noting that the right--hand side of \eqref{rayleigh} is proportional to the curvature $\bU''(z)$, we presume its influence on $w$ and $\tc$ be small, and that successively better approximations are obtained by iterative solutions by method of dominant balance \citep[cf.\ e.g.][]{bender78}. This is our understanding of the 'near--potentiality' assumption. We let $w(z)=w^{(0)}(z)+\varepsilon w^{(1)}(z)+...$ and verify that
\begin{subequations}
\begin{eqnarray}
  w^{(0)}(z)&=& w^{(0)}(0)\frac{\sinh k(z+h)}{\sinh kh},\label{w0}\\ 
  \varepsilon w^{(1)}(z)&=& \frac{w^{(0)}(0)}{k}\int_{-h}^z \frac{ \mathbf{k} \cdot \mathbf{U} ''(\zeta)}{ \mathbf{k} \cdot\Delta \mathbf{U} (\zeta)-k\tilde{c}}\frac{\sinh k(\zeta+h)\sinh k(z-\zeta)}{\sinh kh}\mathrm{d} \zeta.
\end{eqnarray}
\end{subequations}
Inserting the leading approximation of $w(z)$ we obtain $\varepsilon I= \varepsilon\Omega_I+...$ with
\begin{equation} \label{OmI}
  \varepsilon\Omega_I(\tilde{c}) = \frac{2\tilde{c}}{k^2c_0^2}\zint \frac{ \mathbf{k} \cdot \mathbf{U} ''(z) \mathbf{k} \cdot\Delta \mathbf{U} (z)\sinh^2 k(z+h) }{[ \mathbf{k} \cdot\Delta \mathbf{U} (z)-k\tilde{c}]\sinh 2kh}\mathrm{d} z.
\end{equation} 

Now inserting the expansion for $w(z)$ into \eqref{K} gives, after some algebra,  
$ K = 2\delta  + (c_0/\tilde{c})(\varepsilon\Omega_{K1}\delta + \varepsilon\Omega_{K2})$ 
with $\delta $ defined in \eqref{deltaL}, and
\begin{subequations}\label{OmK}
\begin{eqnarray}
  \varepsilon\Omega_{K1}(\tilde{c}) &=& \frac{2\tilde{c}}{kc_0}\zint \frac{ \mathbf{k} \cdot \mathbf{U} ''(z)}{ \mathbf{k} \cdot\Delta \mathbf{U} (z)-k\tilde{c}}\frac{\sinh k(z+h)\sinh kz}{\sinh kh}\mathrm{d} z, \\
  \varepsilon\Omega_{K2}(\tilde{c}) &=& \frac{2\tilde{c}}{k^2c_0^2}\zint \frac{ \mathbf{k} \cdot \mathbf{U} ''(z)}{[ \mathbf{k} \cdot\Delta \mathbf{U} (z)-k\tilde{c}]} \frac{\sinh k(z+h)}{\sinh 2kh}\notag \\
  &&\times\left[\int_z^0 \mathbf{k} \cdot \mathbf{U} '(\zeta) \sinh k(2\zeta+h-z)\mathrm{d} \zeta\right]\mathrm{d} z.
\end{eqnarray}
\end{subequations}
We have introduced a key parameter in our theory, 
\begin{eqnarray}
  \delta ( \mathbf{k} )&=& \int_{-h}^0\frac{ \mathbf{k} \cdot \mathbf{U} '(z)\sinh 2k(z+h)}{kc_0\sinh 2kh}\mathrm{d} z,\label{deltaL}
\end{eqnarray}
which is a dimensionless depth-averaged shear, to which we shall refer frequently.

We thus obtain the approximation for \eqref{fulldispAlt} on the desired form 
\begin{equation} \label{dispDelta}
  \tilde{c}^2+2c_0\tilde{c}\delta -c_0^2+c_0^2\Delta (\tilde{c}) +... = 0
\end{equation} 
with $\Delta (\tilde{c}) = \varepsilon\Omega_I(\tilde{c})+ \varepsilon\Omega_{K1}(\tilde{c})\delta+\varepsilon\Omega_{K2}(\tilde{c})$, which may be simplified with some tedious manipulation (see Appendix \ref{app:Delta}) to the form
\begin{subequations}\label{DeltaL}
\begin{align}
  \Delta (\tilde{c}) =& -\frac{2\tilde{c}}{kc_0^2}\int_{-h}^0\frac{ \mathbf{k} \cdot \mathbf{U} ''(z)}{ \mathbf{k} \cdot\Delta \mathbf{U} (z)-k\tilde{c}}\frac{\sinh k(z+h)\sinh kz}{\sinh kh}\left[  \tilde{U}-\tilde{u}(z) \right]\mathrm{d} z, \\
  \tilde{u}(z) =& -\frac{\sinh kh}{\sinh kz} \int_z^0\frac{2 \mathbf{k} \cdot \mathbf{U} (\zeta)\cosh k(2\zeta+h-z)}{\sinh 2kh}\mathrm{d} \zeta.
\end{align}
\end{subequations}
The weighted depth--averaged velocity, as introduced by \citet{skop87} and used extensively by \citet{kirby89}, is
\be \label{barU}
  \tilde{U}( \mathbf{k} )=\frac{ \mathbf{k} \cdot \mathbf{U} _0}{k}-c_0\delta( \mathbf{k} )  = \int_{-h}^0\frac{2 \mathbf{k} \cdot \mathbf{U} (z)\cosh 2k(z+h)}{\sinh 2kh}\mathrm{d} z
\ee
(the middle form is found by partial integration). Inspection reveals that $\tilde{u}(z)$ is a possible generalization of the depth averaged velocity $\tilde{U}$ when averaging is only carried out down to a depth $z$ rather than the full depth. In particular, $\tilde{u}(-h)=\tilde{U}$ and $\lim_{z\to 0}\tilde{u}(z) =  \mathbf{k} \cdot \mathbf{U} _0/k$.

The effect of shear is now contained in the dimensionless mean shear $\delta$, and the effect of curvature of $\bU(z)$ has successfully isolated in the dimensionless quantity $\Delta(\tc)$. In practice, $\Delta(\tc)$ is calculated using an approximate value of $\tc$, as will be detailed below. If $\kdU=k\tc$ somewhere in the integration range, the outer integral in \eqref{DeltaL} has a pole and the principal value should be taken \citep{shrira93}. Deep water limits of $\delta$ and $\Delta$ are quoted in appendix \ref{app:deep}.

Equation \eqref{dispDelta} now makes explicit what the true small parameter is, namely $\Delta(\tc)$. Note that, like \citet{shrira93}, we find that the effect of the curvature, corresponding to the right--hand side of \eqref{rayleigh}, only need be small in a depth--integrated sense.

\subsection{New approximation for $\tc$}\label{sec:approx}

Solving \eqref{dispDelta} with respect to $\tc$ gives
\be\label{cSq}
  \tc \approx c_0[\sqrt{1+\delta^2-\Delta}-\delta].
\ee
In accordance with the 'near--potentiality' assumption we presume $\Delta\ll 1$ while making no assumptions about the strength of the (non--dimensional) averaged shear. This gives a new approximation 
\be\label{c12}
  c\approx c_\approx + \tilde{c}_{\approx,\text{2nd}}
\ee
with the leading order approximation
\begin{equation} \label{cEL}
  c_\approx( \mathbf{k} ) 
  = \frac{ \mathbf{k} \cdot \mathbf{U} _0}k + c_0\left(\sqrt{1+\delta ^2}-\delta \right) ,
\end{equation} 
which is our main result, and a second order correction
\begin{equation} \label{c2nd}
  \tilde{c}_{\approx,\text{2nd}} = -\frac{c_0\Delta (\tilde{c}_\approx)}{2\sqrt{1+\delta ^2}}.
\end{equation} 
To calculate $\Delta $ in practice, the first order estimate is inserted for $\tilde{c}$, as indicated.
We test the new approximation $c_\approx$ in a range of cases in section \ref{sec:numerics}. 

A few remarks about the approximation \eqref{cEL} are warranted. The first of these is that when $\bU(z)$ is a linear function of $z$, i.e.\ $\bU(z)=\bU_0 + \bU_0' z$, equation \eqref{cEL} gives the \emph{exact} answer which is well known to be $\tilde{c}=c_s$ \citep[e.g.,][]{ellingsen14}, with
\begin{equation}  \label{cs}
  c_s( \mathbf{k} ) =  -\frac{ \mathbf{k} \cdot \mathbf{U} '_0}{2k^2}\tanh kh + \sqrt{c_0^2+\left(\frac{ \mathbf{k} \cdot \mathbf{U} '_0}{2k^2}\tanh kh\right)^2}.
\end{equation} 
(This relation may have been given first by \cite{craik68}). 
The majority of the analytical literature on waves on shear flow concerns this type of flow, for which reason it is an important benchmark. 

We secondly remark that the above derivations were performed for an arbitrary $\bk$. No assumptions were made which might restrict the theory to a particular range of wavelengths. 

Thirdly one may note that the widely used (3D) Kirby \& Chen approximation \eqref{cKC}, which we will review in a moment, is conventionally implemented by calculating $\tilde{U}$ rather than $\delta $. If this is preferred, $c_\approx$ is expressed in terms of $\tilde{U}$ using \eqref{barU}: 
\begin{equation} \label{cELU}
  c_\approx( \mathbf{k} ) = \sqrt{c_0^2 + (\tilde{U}- \mathbf{k} \cdot \mathbf{U} _0/k)^2} + \tilde{U} ,
\end{equation} 
Numerical implementation of \eqref{cEL} and the 3DKC thus involves essentially identical complexity and calculational effort.

Finally, we note that $\tc_\approx$ is positive for all values of $\delta$ as physically it must be.

\subsection{Comparison with the approximation of Skop/Kirby \& Chen}

We observe that in the special case $\delta \ll1$ equation \eqref{cEL} coincides, modulo a term of order $\delta^2$, with 
the 3DKC:
\be \label{cKC}
  c_\text{KC}( \mathbf{k} )=\frac{ \mathbf{k} \cdot \mathbf{U} _0}k + c_0(1-\delta ) = c_0 +\tilde{U}
\ee
The weighted depth averaged velocity $\tilde{U}$ was defined in \eqref{barU}. 
This first order expression (in orders of $kc/\kU$) was derived by \citet{skop87}, and extended to next order by \citet{kirby89}.  
In contrast with approximation \eqref{cEL}, the 3DKC can predict nonsensical, negative values for $\tc$ when $\delta>1$, corresponding to strongly sheared flows. 

It is clear from this, and further discussed below, that the 3DKC works well when $\delta$ is small compared to unity. In certain important cases $\delta( \mathbf{k} ) $ may \emph{not} be small, in which case \eqref{cEL} is superior to \eqref{cKC} except in special cases where cancellations occur, as will be explained in section \ref{sec:error}. Indeed, should $\delta( \mathbf{k} ) >1$, the 3DKC predicts $\tilde{c}<0$ which is physically unacceptable.

From \eqref{cSq} a second order term may also be derived, equal to $c_0(\delta^2-\Delta)/2$. An ``extended KCA'' is then found by adding this term to to \eqref{cKC}, an alternative to the second order expression proposed by \citet{kirby89}. Unlike \eqref{cKC} the ``extended KCA'' is always positive, and is second order accurate when $\delta $ is small, but \eqref{c2nd} is typically superior when $\delta $ approaches unity. We have not studied this approximation in detail.

Compared to the 2nd order approximation of \citet{kirby89}, \eqref{c2nd} has the clear advantage that it makes no assumptions about the length or velocity of waves, nor the strength of the shear, and requires 
the same criterion as the leading approximation to hold, given in \eqref{ELcrit} below, making it well controlled.  
Our examples indicate that the 2nd order expression by \citet{kirby89} is generally robust and accurate, but we have not succeeded in identifying the criteria for this to hold true. \eqref{c2nd} is also arguably a simpler and more transparent expression than the 2nd order correction derived by Kirby \& Chen, and our own experience is that it is significantly easier to implement, while admitting that this is to an extent a point of preference.

\subsection{Applicability criteria and error estimates}\label{sec:error}

Our goal in this section is two--fold. Firstly we derive the pertinent applicability criteria for the new approximation \eqref{cEL} as well as the 3DKC \eqref{cKC}; these are found in Eqs.~\eqref{ELcrit} through \eqref{KCAcrit}. We use these results to explain the surprising success of the 3DKC, not previously understood to our knowledge --- Indeed it seems to us that the sundry applicability ranges at play have been a source of some puzzlement, evident, for example, in the discussions of \cite{swan00}. The applicability of the short--wave approximation due to \citet{shrira93} is discussed in 
Appendix 
\ref{app:shrira}.

We shall see in the following that inclusion of the second order term in \eqref{c12} strongly increases accuracy for all wavelengths in the examples tested. With the exception of two extremely strongly sheared flows considered in section \ref{sec:KCfail} (where, we shall see, applicability criteria are not satisfied), the estimate \eqref{cSq} is accurate to a percent or less. The criterion that $\tc_\approx$ from \eqref{cEL} is a good approximation we then take to be that $(\tc-\tc_\approx)/\tc_\approx \ll 1$ where we use \eqref{cSq} for the ``exact'' value. 
For approximation \eqref{cEL} this means
\be
  |\Delta| \ll 2|\delta\sqrt{1+\delta^2}-1-\delta^2|.
\ee
The right hand side is strictly greater than $1$ for all $\delta$.
A sufficient and far simpler criterion for this to hold is that
\begin{equation} \label{ELcrit}
  |\Delta (\tilde{c})| \ll 1 .
\end{equation} 
Since $|\Delta|$ is a measure of the 
effect of the right--hand side of the Rayleigh equation \eqref{rayleigh} on $\tc$, i.e.\ the effect of the 
curvature of the velocity profile, criterion \eqref{ELcrit} is a formal 'near--potentiality' criterion.

In contrast, a sufficient criterion of applicability of the 3DKC \eqref{cKC} based on the same argument 
is that 
\be\label{KCAcritExtra}
	\left|\frac{\delta^2-\Delta}{2(1-\delta)}\right| \ll 1
\ee
assuming $\delta<1$.
	
A sufficient criterion for this to hold is the \emph{double} criterion
\begin{equation} \label{KCAcrit}
  \delta^2\sim |\Delta (\tilde{c})|\ll 1.
\end{equation}   
Particularly, in cases where 
$|\Delta| \ll 1$ 
is satisfied but $\delta $ is not small, the new approximation \eqref{cEL} is most often superior. However, \eqref{KCAcritExtra} shows that situations exist where the 3DKC happens to be accurate even when $\delta$ is not so small, 
because another sufficient criterion implying \eqref{KCAcritExtra} is that 
\be
  |\delta^2-\Delta|\ll 1, 
\ee
provided $2(1-\delta) \sim \mathcal{O}(1)$.
In such cases a fortunate cancellation occurs in the next order correction to \eqref{cKC}, rendering 
the first order approximation
accurate even when $\delta^2$ and $\Delta$ are not so small respectively, 
as we shall see in numerical examples in Section \ref{sec:numerics}.

\subsubsection{The broad applicability of 3DKC explained}

Remarkably, the 3DKC \eqref{cKC} can be derived by \emph{either} assuming $kc\gg| \mathbf{k} \cdot \mathbf{U} |$ \citep{stewart74,kirby89}\footnote{The assumption $kc\gg| \mathbf{k} \cdot \mathbf{U} |$ is not in general satisfactory since $ \mathbf{U} $ and $c$ (unlike $\Delta  \mathbf{U} $ and $\tilde{c}$) depend on the choice of reference system. In many cases this is easily rectified in a way amounting to the same analysis. } \emph{or} by taking the in some sense opposite limit $k\to\infty$. This surprising coincidence was noted by \citet{skop87}, but seems not to have been realised by either Kirby \& Chen or Stewart \& Joy. We are now in a position to provide an explanation.

Inspecting the definition \eqref{deltaL}, and provided `near potentiality' \eqref{ELcrit} is satisfied, the criterion 
$\delta^2 \ll 1$ from \eqref{KCAcrit} 
implies that the 3DKC is sure to be excellent in three different limiting cases 
where $\delta$ vanishes:
\begin{enumerate}
  \item Weak current, $kc_0/\kU\to\infty$;
  \item When the shear is weak, $ \mathbf{k} \cdot \mathbf{U} '(z)/k^2 c_0\to0$;
  \item When wavelength is short, $k\to\infty$.
\end{enumerate}
That $\delta\to 0$ in the first case may be seen from \eqref{cKC}, whereby $\delta=\kU_0/kc_0-\tU/c_0$. The third follows from the large $k$ asymptotic $\delta \sim \kU_0'/(2k^2 c_0) \sim \kU_0'/(2g^\frac12 k^\frac32)$ which tends to zero in this limit.

\citet{kirby89} derive (the 2D case of) equation \eqref{cKC} based on, essentially, the first of these cases, but conjecture that the true condition might be weak shear. \cite{skop87} remarks that essentially the same approximation may be derived assuming short waves instead. These are exactly the three regimes listed.

\subsubsection{Error estimates}

Noting that the second order approximation constitutes a significant improvement in accuracy for $\tilde{c}$ 
in all examples considered, we propose that its magnitude relative to the first-order approximation of $\tilde{c}$ provides a rough estimate of the error of using \eqref{cEL} and \eqref{cKC}, respectively,
\begin{equation} \label{err}
  \mathrm{err}_\approx \sim \frac{c_0|\Delta (c_\approx)|}{2\tc_\approx\sqrt{1+\delta ^2}}, ~~~
  \mathrm{err}_\text{KC} \sim \frac{|\delta ^2-\Delta (\tilde{c}_\text{KC})|}{2(1-\delta )}.
\end{equation} 

In general, $\mathrm{err}_\approx$ is of order $\Delta $, while $\mathrm{err}_\text{KC}$ is of order $\max(\delta ^2,\Delta )$. However, as mentioned the 3DKC can conspire to be smaller than this due to partial cancellation between $\delta$ and $\Delta/\delta$ ($\mathrm{err}_\text{KC}$ is only valid for $\delta $ smaller than, and not too close to, $1$.)


\subsection{Group velocity}\label{sec:group}

The approximate group velocity according to the approximation \eqref{cEL} is found as
\begin{equation} 
  c_{g,\approx}( \mathbf{k} ) = \frac{\mathrm{d}}{\mathrm{d} k}kc_\approx( \mathbf{k} ) = \frac{ \mathbf{k} \cdot \mathbf{U} _0}k - C_L + (1+\delta ^2)^{-1/2}(c_{g0}+C_L\delta ),
\end{equation} 
where $C_L=\frac{\mathrm{d}}{\mathrm{d} k}(kc_0\delta )$ and $c_{g0}=\frac12 c_0(1+2kh\mathrm{csch} 2kh)$ is the group velocity with no current. Explicitly
\begin{equation} 
  C_L  =(1-2kh\coth 2kh)c_0\delta +2\int_{-h}^0\frac{ \mathbf{k} \cdot \mathbf{U} '(z)(z+h)\cosh 2k(z+h)}{\sinh 2kh}\mathrm{d} z.
\end{equation} 
For reference, the group velocity predicted with 3DKC \eqref{cKC} is  
$c_{g,\text{KC}}= \mathbf{k} \cdot \mathbf{U} _0/k+c_{g0}-C_L$. 
A comprehensive discussion of the group velocity using the 3DKC was recently given by \cite{banihashemi17}.

\section{Application to critical layers}\label{sec:critlayers}

Especially in the presence of strong surface shear, critical layers could occur near enough to the surface to affect wave properties. Critical layers occur when a critical depth $z_c$ exist so that $ \mathbf{k} \cdot \mathbf{U} (z_c)=kc$. 
In the presence of a critical layer the eigenvalues for the phase velocity $c$ acquire an imaginary part whose sign determines the stability of the flow. A thorough background is provided by \citet{drazin81}.  A fully realistic treatment of this problem requires inclusion of viscous effects \citep{velthuizen69}, which is beyond the scope of this study. The inviscid problem is nevertheless 
an informative model and much of the literature concerns stability of the Rayleigh equation \eqref{rayleigh} rather than the full Orr--Sommerfeld problem \citep{morland91,young14}. 

We show in appendix \ref{app:critlayers} how equation \eqref{fulldisp} allows us to generalize the approximate treatment by \citet{shrira93} in a simple way. We conclude, like \citet{shrira93}, that instability is predicted when $ \mathbf{k} \cdot \mathbf{U} ''(z_c)<0$, but derived here under far less restrictive assumptions. The explicit approximation for $\mathrm{Im}(c)$ is 
given in appendix \ref{app:critlayers}. The prediction is made by linearizing with respect to $\Delta/\delta$, and is valid only up to linear order in this parameter.

\section{Numerical results}\label{sec:numerics}

In this section we test our approximation \eqref{cEL} and \eqref{c2nd} for different shear flows, with special emphasis on error estimates and the smallness parameters considered in section \ref{sec:error}. For all flows we have also calculated $\tilde{c}( \mathbf{k} )$ with high accuracy using the piecewise-linear approximation (PLA) detailed in \citet{smeltzer16}, allowing us to accurately compare with the ``exact'' answer for all flows considered. The PLA calculations all have relative accuracy better than $10^{-4}$.

\subsection{Typical wind-driven flows}\label{sec:wind}

\begin{figure}[ht]
  \begin{center}
    \includegraphics[width=\textwidth]{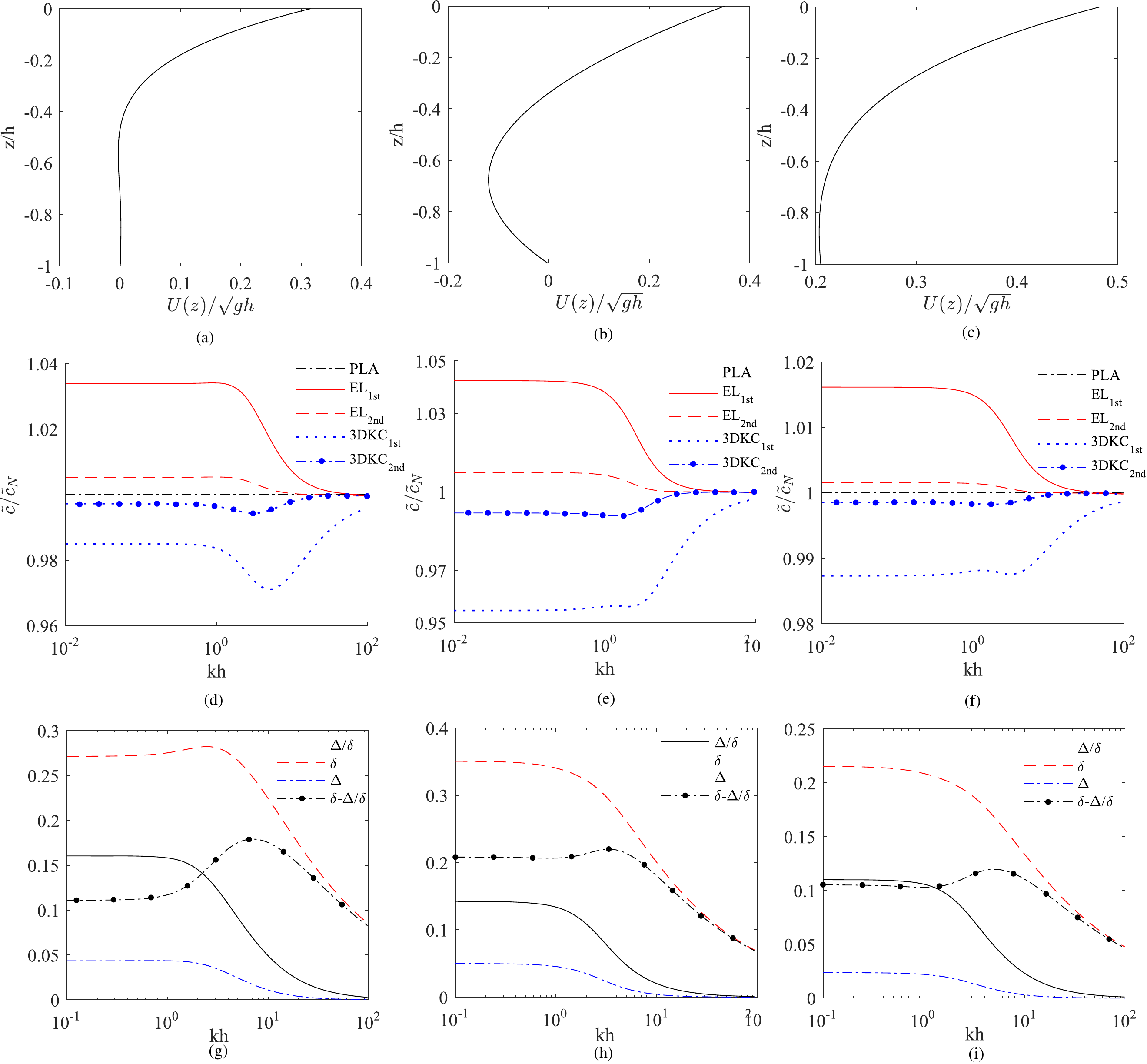}
    \caption{Approximate dispersion relations applied to different wind-drift shear currents \citep{swan00}. Results for the shear profiles in panels (a,b,c) are found in their respective columns. (d,e,f) show calculated estimates of intrinsic velocity $\tilde{c}$ relative to the ``exact'' value calculated with the piecewise--linear method (PLA). Results are calculated for the 1st and 2nd order approximations found herein in Eqs.~\eqref{cEL} and \eqref{c2nd}, respectively, and those due to \citet{kirby89}. Panels (g,h,i) show the parameters relevant to applicability of the approximations. $\Delta (\tilde{c})$ was calculated from the PLA value of $\tilde{c}$.
    }
    \label{fig:wind}
  \end{center}
\end{figure}

Flows with strong shear near the surface will affect the dispersion the most. A typical example is wind-driven flow. Examples of wind driven flow profiles were collected and analysed by \citet{swan00} (see further references therein); we use the velocity profiles from their figure 2, assuming waves propagating downstream. Results are shown in Figure \ref{fig:wind}.

For all of these flows the new approximation \eqref{cEL} and the 3DKC \eqref{cKC} are essentially equally good, which accords with our analysis in section \ref{sec:error} since $\delta $ remains significantly less than $1$ for all $k$. Exactly which of the two happens to be closest to the exact solution for some $k$ depends on the exact shape of $U(z)$, and we do not see a way to predict this short of a full calculation. It is notable that the 3DKC works well even for $\delta $ as large as $0.35$, which is due to the partial cancellation between $\delta ^2$ and $\Delta $, criterion \eqref{KCAcritExtra}. 

Both our 2nd order approximation \eqref{c2nd} and the higher order expression due to \cite{kirby89} improve accuracy, and are essentially equally good for these moderately sheared flows. 

We have performed the same calculation for a shear--assisted wave, equivalent to letting $\bU\to-\bU$ in Fig.~\ref{fig:wind}. The same conclusions hold, both first order expressions are accurate to better than $3\%$, and both 2nd order forms are much better than this. 

\subsection{Comparison with exact cases}

\begin{figure}[ht]
  \begin{center}
    \includegraphics[width=\textwidth]{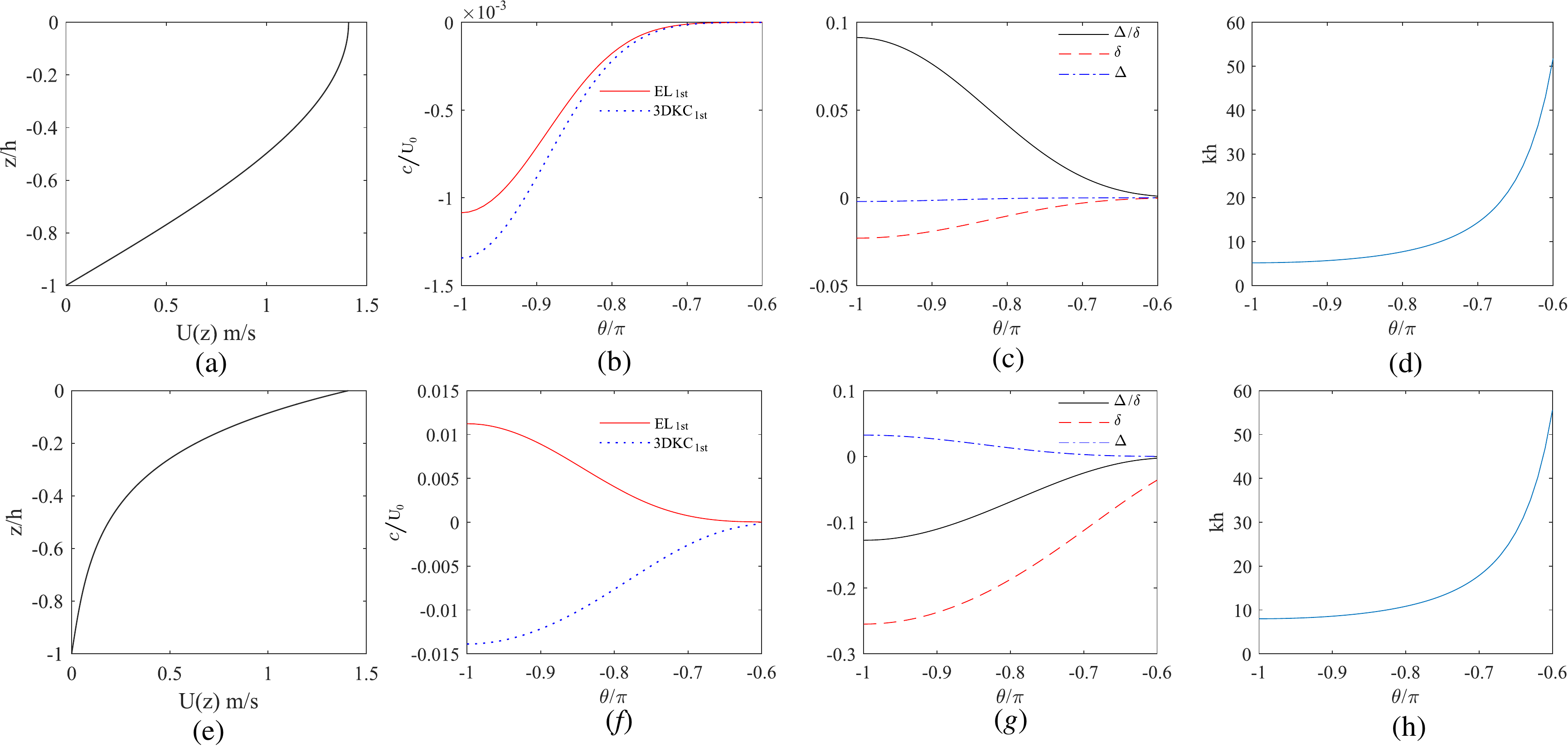}
    \caption{Approximate dispersion relations used for waves with exact phase velocity $c=0$ on profile \eqref{peregrine}, for different propagation directions $\theta$. In all panels $U_0/\sqrt{gh}=0.45$. $h U_0'/U_0=0$ (a-d) and $4$ (e-h); $\kappa$ is determined from the condition $U(-h)=0$. (a,e) show velocity profiles for each row. (b,f) show phase velocities from different approximations; see Fig.~\ref{fig:wind} for abbreviations. (c,g) compares the different smallness parameters considered in section \ref{sec:error}. Wave number $k(\kappa,U_0,U_0')$ is calculated from Eq.~(4.21) of \citet{peregrine76}, and plotted in panels (d,h).  }
    \label{fig:peregrine}
  \end{center}
\end{figure}

We go on to test the different approximations for a particular class of shear currents analysed by \citet{peregrine76}, $ \mathbf{U} (z) = U(z)\mathbf{e}_x$ with
\begin{equation} \label{peregrine}
  U(z)=U_0\cosh \kappa z + U_0' \kappa^{-1}\sinh \kappa z.
\end{equation} 
For the specific case $c=0$, 
the Rayleigh equation \eqref{rayleigh} can be solved exactly giving $w(z)=w(0)\sinh K(z+h)/\sinh K h$ with $K=\sqrt{k^2+\kappa^2}$. The wave number $k$ is a function of the three parameters $\kappa, U_0$ and $U_0'$ and solves an implicit dispersion relation. See \citet[][pp.\ 79--82]{peregrine76} for full details. We choose $U(-h)=0$, which fixes $\kappa$ implicitly.

Results are shown in figure \ref{fig:peregrine}, where two different profiles of type \eqref{peregrine} are analysed 
with $\theta$ being the angle between $\bk$ and $\bU$; velocity profiles are shown in panels (a) and (e), and each row presents calculations pertaining to their respective profiles. The value of $kh$ which corresponds to $c=0$ is shown in panels (d) and (h), showing that the wavelengths become very short as $\theta$ approaches $\pi/2$, whereas the waves near $\theta=-\pi$ the waves 
have wavelength comparable to $h$, and are thus affected by the flow within a significant portion of the water column. 
The small values of $\delta$ for the profile in Fig.~\ref{fig:peregrine}a means the two predictions are almost identical and both predict $c\sim 10^{-3}U_0$. For the profile in Fig.~\ref{fig:peregrine}e with strong surface shear, $\delta$ is higher, yet there is no significant difference in prediction accuracy between the two models, which can be explained by partial cancellation between $\Delta$ and $\delta$ in $\mathrm{err}_\mathrm{KC}$ in Eq.~\eqref{err}.

\subsection{Fully 3 dimensonal flows}

\begin{figure}[ht]
  \begin{center}
    \includegraphics[width=.9\textwidth]{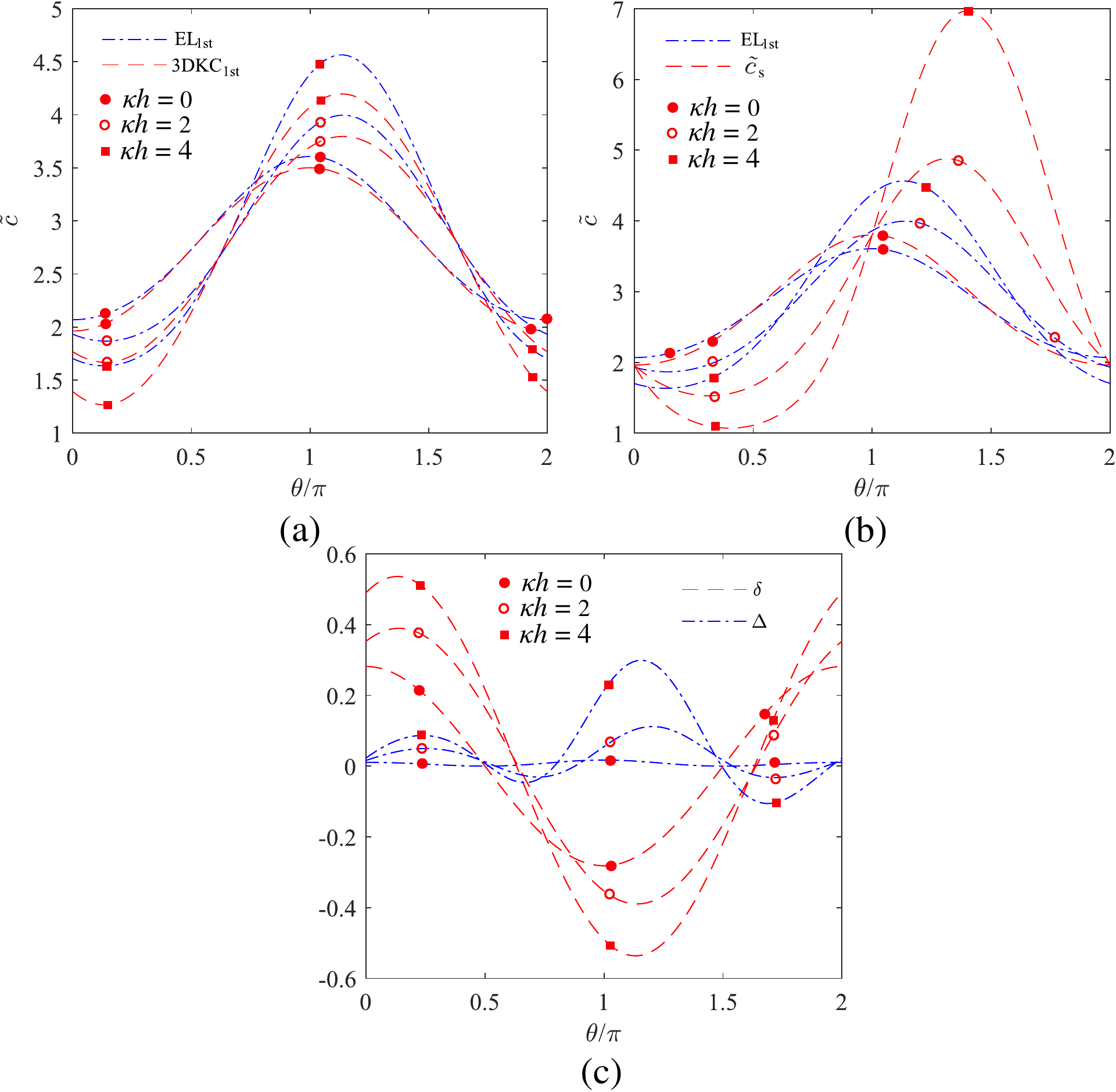}
    \caption{Demonstration of approximate dispersion relation for the direction-changing flow \eqref{dirch} with $U_0/\sqrt{gh}=0.5, \alpha h=1, \phi=0$        and $kh=1$.
    All panels show variation as a function of propagation direction $\theta$ for three different values of 
    $\kappa h$.
   (a) Comparison of \eqref{cEL} and the 3DKC, \eqref{cKC}. (b) Comparison of \eqref{cEL} and the na\"ive estimate $\tilde{c}_s$. (c) Parameters $\delta$ and $\Delta$ as functions of $\theta$.}
    \label{fig:3D}
  \end{center}
\end{figure}

In realistic settings, the direction of $ \mathbf{U} (z)$ may vary with depth. An example could be a wind-driven flow across a tidal current. As a model flow we use an exponential/trigonometric profile
\begin{equation} \label{dirch}
   \mathbf{U} (z) = U_0\sinh\alpha(z+h)[\cos(\kappa z+\phi)\mathbf{e}_x + \sin(\kappa z+\phi)\mathbf{e}_y]
\end{equation} 
where $\alpha$ and $\kappa$ are parameters for the vertical and horizontal depth variation.
Results are shown in Fig.~\ref{fig:3D} for different propagation angles $\theta$ and three different values of the parameter $\kappa h$ ranging from $\kappa h=0$ (no directional variation) to $\kappa h=4$ (strongest directional variation) as shown. We use $kh=1$ everywhere,
and $\theta$ is the angle between $\bk$ and $\bU(0)$.

There are several interesting phenomena to note. Panel (a) shows significant discrepancy between our model \eqref{cEL} and the 3DKC, increasing with stronger vertical shear,
in the vicinity of $\theta=0$. We do not have ``exact'' calculations to compare with for the direction-changing profile, yet the previous error and applicability analysis gives an explanation of discrepancies when the key parameters $\delta$ and $\Delta$ are also considered, shown in Fig.~\ref{fig:3D}c. At the angles where the discrepancy is greatest ($\theta\approx 0.2\pi$), $\delta$ is large while $\Delta$ is too small for a cancellation to occur according to \eqref{KCAcritExtra}. Our analysis strongly indicates that our model \eqref{cEL} should be superior here. In contrast, near $\theta=\pi$ the difference between the two estimates is smaller than might be expected from the high $\delta$ value, as is explained by the partial cancellation 
between $\delta^2$ and $\Delta$ in $\mathrm{err}_\mathrm{KC}$ in Eq.~\eqref{err}.

Finally Fig.~\ref{fig:3D}b shows clearly that the na\"ive estimate for $\tilde{c}$ by extending the surface shear linearly into the deep using $\tilde{c}_s$ [Eq.~\eqref{cs}] will work poorly in this case, where inclusion of the change of direction with depth is important for accurate dispersion estimation.

\subsection{Strongly sheared flows}\label{sec:KCfail}

We finally consider three strongly sheared examples where $\delta $ is so large that the 3DKC fails for a range of wavelengths. Due to finite depth, phase velocities are limited in the long wave limit. Our first two profiles are of finite depth, and their variation over the water column exceeds the fastest possible wave speed, hence waves can never be considered ``fast'' nor the shear weak. The third is a deep water example with strong surface shear. All currents are unidirectional and we choose $ \mathbf{k} $ along the current, an effectively 2D flow which is adequate as a demonstration. For all three example flows we have calculated $\tilde{c}$ to an accuracy better than $10^{-4}$ using the piecewise linear approximation \citep{smeltzer16}, allowing us to compare approximations to the $\tilde{c}(k)$ which at the relevant level of accuracy can be considered ``exact''. 
In our examples we consider waves travelling downstream.

\begin{figure}
  \includegraphics[width=\textwidth]{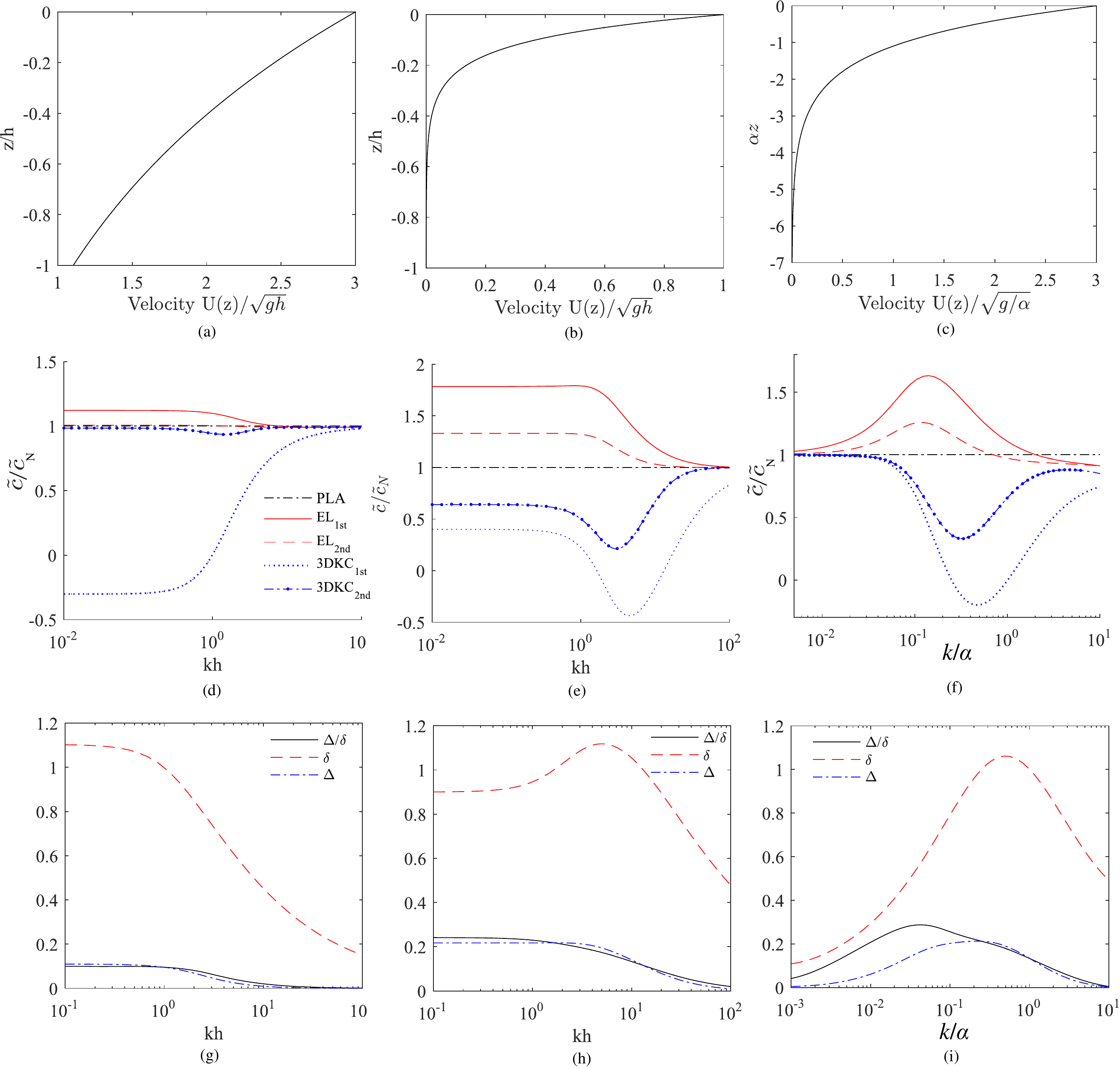}  
  \caption{Comparison of different approximation models for three strongly sheared velocity profiles. (a,d,g) pertain to $U_1(z)$ from eq.~\eqref{profile2},  (b,e,h) to $U_2(z)$ from eq.~\eqref{profile1}, and (c,f,i) to $U_3(z)$ from \eqref{profile3}. (a,b,c): Velocity profiles. (d,e,f): 1st and 2nd order estimates using the present model to first and second order, respectively EL$_\text{1st}$ [Eq.~\eqref{cEL}], and EL$_{2nd}$ [Eq.~\eqref{c2nd}], as well as the 1st and 2nd order approximations of \citet{kirby89} (3DKC$_{1st}$ and 3DKC$_{2nd}$, respectively), relative to the high accuracy calculation with the piecewise linear approximation (PLA). 
  The legend in panel (d) applies also to (e) and (f). 
  (g,h) and (i): 
  applicability parameters for the models as discussed in Section \ref{sec:error}.
  }
  \label{fig:extreme}
\end{figure}

The first profile we consider is an example where our approximation \eqref{cEL} is successful whereas the 1st order 3DKC \eqref{cKC} fails completely:
\begin{equation} \label{profile2}
  U_1(z) = 3\sqrt{gh}\exp(z/h).
\end{equation} 
This profile is considered in panels a,d and g of figure \ref{fig:extreme}. $U_1(z)$ has strong shear, and its variation $|\Delta U_1|$ exceeds $\tilde{c}$ for a large part of the wave number spectrum. The velocity variation over the column is not unreasonable for a region of rapid flow, for example over a local shallow in a river --- a variation $U_0-U(-h)\sim 2\sqrt{gh}$ as here is obtained, for instance, for flow of surface velocity $4$m/s in $40$cm depth --- or it might describe a fast film flow of depth $1$cm with surface velocity $60$cm/s. 

Although the 3DKC fails because $\delta $ is too large, the 'near-potentiality' criterion \eqref{ELcrit} is very well satisfied, and $c_\approx$ is within a few \% of the correct value. 
Figure \ref{fig:extreme}a,d,g demonstrate the improved robustness of our approximation \eqref{cEL} compared to the 
Skop/Kirby \& Chen model. 
The 3DKC becomes negative for long wavelengths, which is to say phase velocity should change sign in a system following the surface, which is physically unacceptable: the dispersion relation for a linear shear current always has one upstream and one downstream solution however strong the shear. For any smooth, non-linear velocity profile, there always exists an even more strongly sheared linearly varying current, and since $\tilde{c}_s$ from \eqref{cs} is always positive, so must $\tilde{c}$.

Insight into the nature of second order correction terms is also offered by figure 
\ref{fig:extreme}d. 
The 2nd order correction to $c_\approx$, equation \eqref{c2nd}, is accurate to $1\%$ or better for all $k$, while the 2nd order term as calculated from the expressions of \citet{kirby89}, $\tilde{c}_\text{KC,2nd}$, also rectifies the deficiencies of its 1st order companion (the expression for $\tilde{c}_\text{KC,2nd}$ is somewhat bulky so is not quoted here) and works well; it has much higher relative error than \eqref{c2nd}, but this in itself is unlikely to be of practical importance. We are at present unable to explain the success of the 2nd order Kirby \& Chen approximation in this case, and thus cannot determine under what conditions this holds true more generally. The criterion for the 2nd order approximation \eqref{c2nd}, on the other hand, is well controlled by condition \eqref{ELcrit}. 


We next consider a particularly difficult case where both shear and curvature are very large, 
\begin{equation} \label{profile1}
  U_2(z) = \sqrt{gh}\exp(10 z/h).
\end{equation} 
$|\Delta U_2|$ exceeds $\tilde{c}$ for much of the water column. $U_2(z)$ is considered in figure \ref{fig:extreme}, panels b,e and h. 
Such a strongly sheared flow could occur locally, and 
might represent a realistic surface jet due to discharge 
of a fast flow 
into still waters, for example a jet speed of about $3$m/s on a $1$m depth. 

We notice that for $U_2(z)$ the 
parameter 
$\Delta $
is about 
$0.22$
for long waves, 
enough for the flow not to 
satisfy the `near-potentiality' criterion \eqref{ELcrit}. No approximation scheme based on near-potentiality can expect to fully succeed in this case. Since $\delta$ exceeds unity, 
the 3DKC once again fails. 
Also $c_\approx$ from \eqref{cEL} is inaccurate in this particular case, overestimating $\tilde{c}$ by about $60\%$ for long waves, yet the result is at least physically meaningful since the correct sign of $\tilde{c}$ is guaranteed. 
For reference, $\tc$ tends to the maximum $\approx 0.5\sqrt{kh}$ as $k\to 0$.
The 2nd order corrected approximation \eqref{c2nd} moreover improves the estimates to the $20\%$ level or better, whereas the 2nd order approximation due to \citet{kirby89} is too small by 
about $40\%$ for long waves and fares poorly for medium waves with $kh\sim 5$ where $\tilde{c}_\text{KC,2nd}$ is only about $20\%$ of the real $\tilde{c}$. This particularly difficult example thus demonstrates the improved robustness of \eqref{cEL} and \eqref{c2nd}. 

We finally consider another difficult case for any 'near-potentiality' approximation scheme, with strong surface shear in deep water,
\begin{equation} \label{profile3}
  U_3(z) = 3\sqrt{g/\alpha}\exp(\alpha z).
\end{equation} 
Since the shear is found near the surface only, the downstream phase velocity is unbounded for long waves, unlike for a linearly varying deep water current. 
Like in the previous cases, a region of wavelengths exists where the 3DKC yields an unphysical negative prediction, $k/\alpha$ between approximately $0.2$ and $1$. The robustness of our new approximation is thus demonstrated once again. It should be noted 
that while the 3DKC fails for medium wavelengths, it 
fares better than $c_\approx$ and even $c_{\approx,\text{2nd}}$ for $k/\alpha \lesssim 0.05$, despite the fact that $\delta$ is as high as $0.6$. Again this is because 
$\Delta/\delta\approx \delta$ here, making 
criterion \eqref{KCAcritExtra} 
well satisfied nevertheless, 
as seen in Fig.~\ref{fig:extreme}i.
This seems like a lucky coincidence, but given similar observations for the wind-driven profiles in section \eqref{sec:wind}, 
it seems indicated that to some extent
it holds more generally for a class of shear currents, a question which should be looked into in the future.


\section{Conclusions}\label{sec:conclusions}

We have presented a comprehensive theory of approximate dispersion relations for linear waves propagating atop a sub-surface horizontal shear current whose magnitude and direction can vary arbitrarily as a function of depth. We present a new analytical approximation which is shown to be more robust than the (3D generalization of the) famous and widely used approximation by 
\citet{skop87} and 
\citet{kirby89}, the 3DKC. The conditions of applicability of our new model are shown to be less restrictive, making it accurate in several realistic situation where the 3DKC is inaccurate or even breaks down. For the cases when the criteria for the 3DKC to work are satisfied, the two models are equally good,
and we do not see a way to predict which approximation happens to give the most accurate result in a given situation short of performing a more accurate calculation at greater cost. 
Our approximations are tested for a range of different example flows.

A 2nd order accurate expression in terms of the suitable small parameter is also derived, and shown to 
greatly 
improve accuracy. The inclusion of the next order may not be called for in many situations, but constitutes a significant improvement in the more difficult situations discussed. The next order correction is robust and has the same condition of applicability as the first order (\citet{kirby89} also derive a 2nd order correction to their model, whose conditions of applicability are not known at present).

Our thorough perturbation analysis is able to explain, for the first time, the success of the 
3DKC 
for many cases,
including situations where the assumptions behind its original derivation are strongly violated. 
Indeed, careful criteria are derived for our approximation, as well as that of Kirby \& Chen, to be accurate.

To leading order our new approximation involves essentially identical computational effort and complexity as that of Kirby \& Chen. Our experience, however, is that our new 2nd order correction is significantly less complicated to implement than its sibling derived by said authors (while admitting that this could be a point of preference), and arguably more physically transparent.

\acknowledgments
  We are endebted to Benjamin K.~Smeltzer and Peter Maxwell for performing the piecewise-linear calculations 
  in Sec.~\ref{sec:numerics}.
  YL is funded by the Faculty of Engineering, Norwegian University of Science and Technology. S{\AA}E acknowledges funding from the Norwegian Research Council (FRINATEK), project number 249740. 
  No new data was generated for the research reported herein, and all equations necessary to reproduce the results are included.

\appendix


\section{Further discussion and mathematical details}

\subsection{Infinite depth expressions}\label{app:deep}
\newcommand{\zintinf }{\int_{-\infty}^0}
We here list key quantities in the limit $kh\to\infty$. Now $c_0=\sqrt{g/k}$, and 
\begin{equation} 
  \delta  = \int_{-\infty}^0\frac{ \mathbf{k} \cdot \mathbf{U} '(z)}{kc_0}\mathrm{e}^{2kz}\mathrm{d} z; ~~~~ 
  \tilde{U} = 2\int_{-\infty}^0  \mathbf{k} \cdot \mathbf{U} (z)\mathrm{e}^{2kz}\mathrm{d} z.
\end{equation} 
The relation $k\tilde{U}= \mathbf{k} \cdot \mathbf{U} _0-kc_0\delta $ still holds. The expression for $\Delta (\tilde{c})$ simplifies greatly, to
\begin{eqnarray}
  \Delta (\tilde{c}) &=& -\frac{\tilde{c}}{kc_0^2}\int_{-\infty}^0\frac{ \mathbf{k} \cdot \mathbf{U} ''(z)}{ \mathbf{k} \cdot\Delta \mathbf{U} (z)-k\tilde{c}}[\tilde{U} - \tilde{u}(z)]\mathrm{e}^{2kz}\mathrm{d} z;\\
  \tilde{u}(z) &=& 2\int_{-\infty}^z  \mathbf{k} \cdot \mathbf{U} (\zeta)\mathrm{e}^{2k(\zeta-z)}\mathrm{d} \zeta.
\end{eqnarray}
The first two solutions to $w(z)$ are
\begin{equation} 
  w^{(0)}(z)=w^{(0)}(0)\mathrm{e}^{kz}; ~~ w^{(1)}(z)=\frac{w^{(0)}(0)}{k}\int_{-\infty}^z\frac{ \mathbf{k} \cdot \mathbf{U} ''(\zeta)}{ \mathbf{k} \cdot\Delta \mathbf{U} (\zeta)-k\tilde{c}}\mathrm{e}^{k\zeta}\sinh k(z-\zeta)\mathrm{d} \zeta.
\end{equation} 
For the purposes of the approximations for group velocity, 
\begin{equation} 
  C_L = \int_{-\infty}^0  \mathbf{k} \cdot \mathbf{U} '(z)(1+2kz)\mathrm{e}^{2kz}/k.
\end{equation} 

\subsection{Simplification of $\Delta(\tc)$}\label{app:Delta}

Here follow the details of the simplification of $\Delta(\tc)=\varepsilon\Omega_I+\varepsilon\Omega_{K1}\delta+\varepsilon\Omega_{K1}$ in Eqs \eqref{OmI} and \eqref{OmK} to the form \eqref{DeltaL}. Using $\delta=(\kU_0-k\tU)/kc_0$ from \eqref{cKC} and $\sinh kh=2\cosh kh/\sinh 2kh$, we write
\be
  \delta \varepsilon \Omega_{K1} = \frac{2\tc}{k^2c_0^2}\zint \Upsilon(z) \frac{\sinh k(z+h)}{\sinh 2kh}2\sinh kz \cosh kh[\kU_0-k\tU]\rmd z
\ee
where the shorthand $\Upsilon(z) = \kU''(z)/(\kdU-k\tc)$ is introduced. Next we perform a partial integration of the inner integral in $\varepsilon\Omega_{K2}$ to obtain
\begin{eqnarray}
  \varepsilon\Omega_{K2}&=& \frac{2\tc}{k^2c_0^2}\zint \Upsilon(z) \frac{\sinh k(z+h)}{\sinh 2kh}\Bigl[\kU_0\sinh k(h-z)-\kU(z)\sinh k(h+z)\notag \\
  &&-2k\int_z^0\kU(\zeta)\cosh k(2\zeta+h-z)\rmd\zeta\Bigr] \rmd z .
\end{eqnarray}
Using that $\sinh k(h-z)-\sinh k(h+z) +2\sinh kz \cosh kh=0$, we find
\begin{eqnarray}
  \Delta &=& -\frac{2\tc}{kc_0^2}\zint \Upsilon(z)\frac{2\sinh k(z+h)}{\sinh 2kh}\Bigl[\tU\sinh kz\cosh kh\notag \\
  &&+ \int_z^0\kU(\zeta) \cosh k(2\zeta+h-z)\rmd\zeta\Bigr]\rmd z,
\end{eqnarray}
which takes the form \eqref{DeltaL} with minimal further manipulation.

\subsection{The short-wave approximation of Shrira}\label{app:shrira}

We consider the approximation due to \citet{shrira93} derived for short waves. 
For large $k$ the integral is dominated by  $|z|\lesssim 1/2k$, and we now assume the wave short enough that a near--surface Taylor expansion $\bU(z)\approx \bU_0 + \bU_0' z + ...$ is in order. 
Terms proportional to $ \mathbf{U} ''(z)$ are then small, and the integral term $I$ [see Eq.~\eqref{Iexp}] is a small correction in equation \eqref{fulldisp}. 
Since 
$\Delta \mathbf{U} \approx  \mathbf{U} '_0 z$ 
we have $ \mathbf{k} \cdot\Delta \mathbf{U} \lesssim  \mathbf{U} '_0/2k$ which is negligible compared to $k\tilde{c}\sim \sqrt{gk}$. We thus obtain $I\approx -(2c_0/\tilde{c})\delta_S$ with the short-wave smallness parameter
\begin{equation} 
  \delta_S = \int_{-h}^0 \frac{ \mathbf{k} \cdot \mathbf{U} ''(z)\sinh^2k(z+h)}{k^2c_0\sinh 2kh}\mathrm{d} z
\end{equation} 
which is a depth-averaged velocity profile curvature. 
Equation \eqref{fulldisp} becomes approximately
\[
  \tilde{c}^2-2\tilde{c} c_0\delta_S+k^{-2}\tilde{c} \mathbf{k} \cdot \mathbf{U} '_0\tanh kh-c_0^2=0.
\]
Assuming $\delta_S$ to be small and expanding to first order gives
\begin{equation} \label{shrira}
  \tilde{c} \approx c_s\left(1+c_0\delta_S/\sqrt{\cdots} \right)
\end{equation} 
where $\sqrt{\cdots}$ is the square root term in \eqref{cs}. This is the leading order of the approximation of \citet{shrira93}. 

The criterion of applicability, $\delta_S\ll 1$, only in general holds for large $k$. For longer waves a cancellation occurs in \eqref{fulldisp}, because, by partial integration, 
\[
  \delta_S =  \mathbf{k} \cdot \mathbf{U} _0'\tanh kh/(2k^2c_0) - \delta .
\]
For short waves both terms on the right hand side are small. For longer waves, the first term no longer is, but it is cancelled by another term in \eqref{fulldisp}. If $\delta $ is also small for long waves, this cancellation effectively replaces the small parameter $\delta_S$ for short waves, by $\delta $ for longer waves. Approximation \eqref{shrira} has no such cancellation, and therefore fails for longer waves, while approximations \eqref{cEL} and \eqref{cKC} typically do not.

\subsection{Stability of a critical layer}\label{app:critlayers}

Let there be a critical layer at $z=z_c$, so that near this depth, $ \mathbf{k} \cdot \mathbf{U} (z)\approx kc+(z-z_c) \mathbf{k} \cdot \mathbf{U} '_c$, where $ \mathbf{U} _c= \mathbf{U} (z_c)=c$ and $ \mathbf{U} '_c= \mathbf{U} '(z_c)$. 
The integral in \eqref{fulldisp} now has a 
pole on the axis of integration. 
It becomes well defined when considered as
the $t\to\infty$ limit of a corresponding initial value problem \citep{peregrine76}. Suppose the wave has been made by a wave paddle with frequency $\omega=kc$, and whose amplitude has increased slowly from zero at $t=-\infty$. The generated wave will have time dependence $\exp(-i\omega t + \epsilon t)$ with $\epsilon = 0^+$. The effect is to replace $kc\to kc+\mathrm{i}\epsilon$, moving the pole slightly off the $z$ axis. 
Approximating $w\approx w^{(0)}$ from \eqref{w0} and using the Sokhotski-Plemelj theorem, a 'near--potentiality'
estimate of $I$ from \eqref{Iexp} is 
\begin{equation} \label{Iim}
  I\approx \mathcal{P}\int_{-h}^0\frac{2 \mathbf{k} \cdot \mathbf{U} '' \sinh^2 k(z+h)\mathrm{d} z}{k( \mathbf{k} \cdot\Delta \mathbf{U} -k\tilde{c})\sinh 2kh}+\frac{2\pi\mathrm{i} \mathbf{k} \cdot \mathbf{U} ''_c\sinh^2k(z_c+h)}{k| \mathbf{k} \cdot \mathbf{U} _c|\sinh 2kh}\equiv I_r+\mathrm{i} I_i.
\end{equation} 
$\mathcal{P}$ denotes the principal value. 
With this, the first-order version of \eqref{dispDelta} becomes $\tilde{c}^2+2c_0\tilde{c}\delta-c_0^2+\mathrm{i} \tilde{c}^2 I_i=0$.
Now suppose the complex $\tilde{c}$ is 
$\tilde{c}=\tilde{c}_r+\mathrm{i} c_i$, and noting that $c_i$ is order $ I_i$, we solve for the real and imaginary parts of the resulting equation to order $ I_i$. 
This results in $\tilde{c}_r\approx \tilde{c}$, and 
\begin{equation} 
  c_i\approx-\tilde{c}^2_\approx I_i/(2c_0 \sqrt{1+\delta^2}).
\end{equation} 
Here, $\tilde{c}_r$ is approximated using, e.g., \eqref{cEL}. 
For the case of short waves in deep water this agrees with the result of \citet{shrira93}. 
Now, $c_i>0$ implies unstable flow. Since $c_0>0$, instability is predicted when $ \mathbf{k} \cdot \mathbf{U} ''_c<0$.




\listofchanges

\end{document}